\documentclass{PoS}

\usepackage{amsmath,amssymb,amsopn}
\usepackage{graphicx}

\newcommand{\be}{\begin{equation}}
\newcommand{\ee}{\end{equation}}
\newcommand{\ba}{\begin{eqnarray}}
\newcommand{\ea}{\end{eqnarray}}
\newcommand{\ban}{\begin{eqnarray*}}
\newcommand{\ean}{\end{eqnarray*}}
\newcommand \nn {\nonumber}

\title{Parton Energy Loss in the Extremely Prolate Quark-Gluon Plasma}

\ShortTitle{Parton Energy Loss in the Extremely Prolate QGP}

\author{Margaret E. Carrington\\
     \\ Department of Physics, Brandon University, Brandon, Manitoba, R7A 6A9 Canada
      \email{carrington@brandon.ca}}

\author{\speaker{Katarzyna Deja}\\
       \\National Center for Nuclear Research, 00-681 Warsaw, Poland
        \email{katarzyna.deja@fuw.edu.pl}}

\author{Stanis\l aw Mr\' owczy\' nski\\
       \\Institute of Physics, Jan Kochanowski University, 25-406 Kielce, Poland 
       \\and National Center for Nuclear Research, 00-681 Warsaw, Poland
       \\ \email{mrow@fuw.edu.pl}}   

\abstract {The energy loss per unit path length of a highly energetic parton scattering elastically in a weakly coupled quark-gluon plasma is studied as an initial value problem. The approach is designed to study unstable plasmas but in the case of an equilibrium plasma the well known result is reproduced. As an example of an unstable system, an extremely prolate plasma, where the momentum distribution is infinitely elongated along one direction, is considered here. The energy loss is shown to be strongly time and directionally dependent and its magnitude can much exceed the energy loss in equilibrium plasma.}

\FullConference{Xth Quark Confinement and the Hadron Spectrum,\\
		October 8-12, 2012\\
		TUM Campus Garching, Munich, Germany}

\begin{document}

\section{Introduction}

When a highly energetic parton travels through a quark-gluon plasma (QGP), it losses its energy due to, in particular, elastic interactions with plasma constituents. This is called {\em collisional energy loss} and was computed for equilibrium QGP twenty years ago, see the review \cite{Peigne:2008wu} and the handbook \cite{lebellac}. The quark-gluon plasma produced in relativistic heavy-ion collisions, however, reaches a state of local equilibrium only after a short but finite time interval, and during this period the momentum distribution of plasma partons is anisotropic. It is important to note that a plasma with an anisotropic momentum distribution is unstable (for a review see \cite{Mrowczynski:2005ki}). Collisional energy loss has been computed for an anisotropic QGP in Ref. \cite{Romatschke:2004au}, but the fact that unstable systems are explicitly time dependent as unstable modes exponentially grow in time was not taken into account. 

We have developed an approach, see \cite{Carrington:2011uj,Carrington:2012hv} for a preliminary account, where energy loss is studied as an initial value problem. The approach is applicable to plasma systems evolving quickly in time. We compute the energy loss by treating the parton as an energetic classical particle with ${\rm SU}(N_c)$ color charge. For an equilibrium plasma the known result is recovered and for an unstable plasma the energy loss is shown to have contributions which exponentially grow in time. In Refs. \cite{Carrington:2011uj,Carrington:2012hv} we have calculated the energy loss in a two-stream system which is unstable due to longitudinal chromoelectric modes and found that it manifests strong time and directional dependence.  In this paper we focus on an extremely prolate quark-gluon plasma with momentum distribution infinitely elongated in one (beam) direction. Such a system is unstable due to transverse chromomagnetic modes and the spectrum of collective excitations can be obtained in explicit analytic form. The system has thus nontrivial dynamics but the computation of energy loss is relatively simple.  After a brief presentation of our approach, we show some of our results. The energy loss grows exponentially and after some time its magnitude is much bigger than in equilibrium plasma.

\section{Formalism}
\label{sec-formalism}

Our approach is classical and thus we start with the Wong equations \cite{Wong:1970fu} describing the motion of classical parton in a chromodynamic field.

\subsection{General energy-loss formula}
\label{subsec-general}

The Wong equations \cite{Wong:1970fu} read
\ba
\label{EOM-1a}
\frac{d x^\mu(\tau)}{d \tau} &=& u^\mu(\tau ) ,
\\
\label{EOM-1b}
\frac{d p^\mu(\tau)}{d \tau} &=& g Q^a(\tau ) \, F_a^{\mu \nu}\big(x(\tau )\big)
\, u_\nu(\tau ) ,
\\
\label{EOM-1c}
\frac{d Q_a(\tau)}{d \tau} &=& - g f^{abc} u_\mu (\tau ) \,
A^\mu _b \big(x(\tau )\big) \,
Q_c(\tau) ,
\ea
where $\tau$, $x^\mu(\tau )$, $u^\mu(\tau)$ and  $p^\mu(\tau)$ are, respectively, the parton's  proper time, its trajectory, four-velocity and  four-momentum; $F_a^{\mu \nu}$ and $A_a^\mu$ denote the chromodynamic field strength tensor and four-potential along the parton's trajectory and $Q^a$ is the classical color charge of the parton; $g$ is the coupling constant and $\alpha_s \equiv g^2/4\pi$ is assumed to be small.  We also assume that the potential vanishes along the parton's trajectory  {\it i.e.} our gauge condition is $u_\mu (\tau ) \, A^\mu _a \big(x(\tau )\big) = 0 $. Due to Eq.~(\ref{EOM-1c}) the classical parton's  charge $Q_c(\tau)$ is a constant of motion within the chosen gauge.

The energy loss is given directly by Eq. (\ref{EOM-1b}) with $\mu = 0$. Using the time $t=\gamma\tau$ instead of the proper time $\tau$ and replacing the strength tensor $F_a^{\mu \nu}$ by the chromoelectric ${\bf E}_a(t,{\bf r})$ and chromomagnetic ${\bf B}_a(t,{\bf r})$ fields, Eq.~(\ref{EOM-1b}) gives
\be
\label{e-loss-1}
\frac{dE(t)}{dt} = g Q^a {\bf E}_a(t,{\bf r}(t)) \cdot {\bf v} ,
\ee
where ${\bf v}$ is the parton's velocity. Since we consider a parton which is very energetic,  ${\bf v}$ is assumed to be constant and ${\bf v}^2 =1$, but the parton's momentum and energy vary. 

Since we deal with an initial value problem, we apply to the field and current not the usual Fourier transformation but the {\it one-sided Fourier transformation} defined as
\ba
\label{1side}
f(\omega,{\bf k}) &=& \int_0^\infty dt \int d^3r
e^{i(\omega t - {\bf k}\cdot {\bf r})}
f(t,{\bf r}) , \\
\label{1side-inv}
f(t,{\bf r}) &=& \int_{-\infty+i\sigma}^{{\infty+i\sigma}} \frac{d\omega}{2\pi} \int \frac{d^3k}{(2\pi)^3}
e^{-i(\omega t - {\bf k}\cdot {\bf r})}
f(\omega,{\bf k}) ,
\ea
where the real parameter $\sigma > 0$ is chosen is such a way that the integral over $\omega$ is taken along a straight line in the complex $\omega-$plane, parallel to the real axis, above all singularities of $f(\omega,{\bf k})$. Introducing the current generated by the parton ${\bf j}_a(t,{\bf r}) = g Q^a {\bf v} \delta^{(3)}({\bf r} - {\bf v}t)$, and using Eqs. (\ref{1side}, \ref{1side-inv}),  Eq.~(\ref{e-loss-1}) can be rewritten:
\be
\label{e-loss-3}
\frac{dE(t)}{dt} = g Q^a
\int_{-\infty +i\sigma}^{\infty +i\sigma}
{d\omega \over 2\pi}
\int {d^3k \over (2\pi)^3}
e^{-i(\omega - \bar\omega)t} \; {\bf E}_a(\omega,{\bf k}) \cdot {\bf v} ,
\ee
where $\bar\omega \equiv {\bf k} \cdot {\bf v}$. 

The next step is to compute the chromoelectric field ${\bf E}_a$. Applying the one-sided Fourier transformation to the linearized Yang-Mills equations, which represent QCD in the Hard Loop approximation, we get the chromoelectric field given as
\be
\label{E-field-k}
E^i_a(\omega, {\bf k}) = -i
(\Sigma^{-1})^{ij}(\omega,{\bf k})
\Big[ \omega j_a^j(\omega,{\bf k})
+ \epsilon^{jkl} k^k B_{0a}^l ({\bf k})
- \omega D_{0a}^j ({\bf k}) \Big] ,
\ee
where  ${\bf B}_0$ and ${\bf D}_0$ are the initial values of the chromomagnetic field and  the chromoelectric induction, and $D^i_a(\omega, {\bf k}) = \varepsilon^{ij}(\omega, {\bf k}) E^j_a(\omega, {\bf k})$
with  $\varepsilon^{ij}(\omega, {\bf k})$ being chromodielectric tensor which equals
\begin{equation*}
\varepsilon^{ij} (\omega,{\bf k})
=  \delta^{ij} + 
{g^2 \over 2 \omega} \int {d^3 p \over (2\pi )^3}
{ v^i \over \omega - {\bf k} \cdot {\bf v} + i0^+} 
{\partial f({\bf p}) \over \partial p^k} 
\Big[ \Big( 1 - {{\bf k} \cdot {\bf v} \over \omega} \Big) \delta^{kj}
+ {k^k v^j \over \omega} \Big] ,
\end{equation*}
where $ f({\bf p})$ is the momentum distribution of plasma constituents. The color indices $a,b$ 
are dropped as $\varepsilon (\omega, {\bf k})$ is a unit matrix in color space. The matrix 
$\Sigma^{ij}(\omega,{\bf k})$ in Eq.~(\ref{E-field-k}) is defined
\be
\label{matrix-sigma}
\Sigma^{ij}(\omega,{\bf k}) \equiv
- {\bf k}^2 \delta^{ij} + k^ik^j
+ \omega^2 \varepsilon^{ij}(\omega,{\bf k}) .
\ee

Substituting the expression (\ref{E-field-k}) into Eq.~(\ref{e-loss-3}), we get the formula
\ba
\label{e-loss-6}
\frac{dE(t)}{dt} &=& g Q^a v^i \int_{-\infty +i\sigma}^{\infty +i\sigma}
{d\omega \over 2\pi i}
\int {d^3k \over (2\pi)^3}
e^{-i(\omega -\bar{\omega})t}
(\Sigma^{-1})^{ij}(\omega,{\bf k})
\\ [2mm]\nn
&\times&
\Big[ 
\frac{i \omega g Q^a v^j}{\omega - \bar{\omega}}
+ \epsilon^{jkl} k^k B_{0a}^l ({\bf k})
- \omega D_{0a}^j ({\bf k}) \Big] .
\ea
As seen, the integral over $\omega$ is controlled by the poles of the matrix $\Sigma^{-1}(\omega,{\bf k})$  
which represent the collective modes of the system. 

\subsection{Equilibrium plasma}
\label{subsec-equilibrium}

When the plasma is stable, all modes are damped and the poles of $\Sigma^{-1}(\omega,{\bf k})$ are located in the lower half-plane of complex $\omega$. Consequently, the contributions to the energy loss corresponding to the poles of $\Sigma^{-1}(\omega,{\bf k})$ exponentially decay in time. The only stationary contribution is  given by the pole $\omega = \bar{\omega} \equiv {\bf k}\cdot {\bf v}$. Therefore, the terms in Eq.~(\ref{e-loss-6}), which depend on the initial values of the fields, are neglected and Eq.~(\ref{e-loss-6}) provides
\ba
\label{e-loss-isotropic}
\frac{d E}{dt} = -i g^2 C_R 
\int {d^3k \over (2\pi)^3} \;
\frac{\bar{\omega}}{{\bf k}^2} \;
\bigg[
\frac{1}
{\varepsilon_L(\bar{\omega},{\bf k})}
+ \frac{{\bf k}^2{\bf v}^2 - \bar{\omega}^2}
{\bar{\omega}^2
\varepsilon_T(\bar{\omega},{\bf k})-{\bf k}^2}
\bigg] ,
\ea
where the color factor $C_R$, which equals $ \frac{N_c^2-1}{2N_c}$ for a quark and $N_c$ for a gluon, results from the averaging over colors of the test parton. The formula (\ref{e-loss-isotropic}) agrees with the standard energy loss due to soft collisions in equilibrium QGP \cite{lebellac}.

\begin{figure}[t]
\centering
\includegraphics[width=0.70\textwidth]{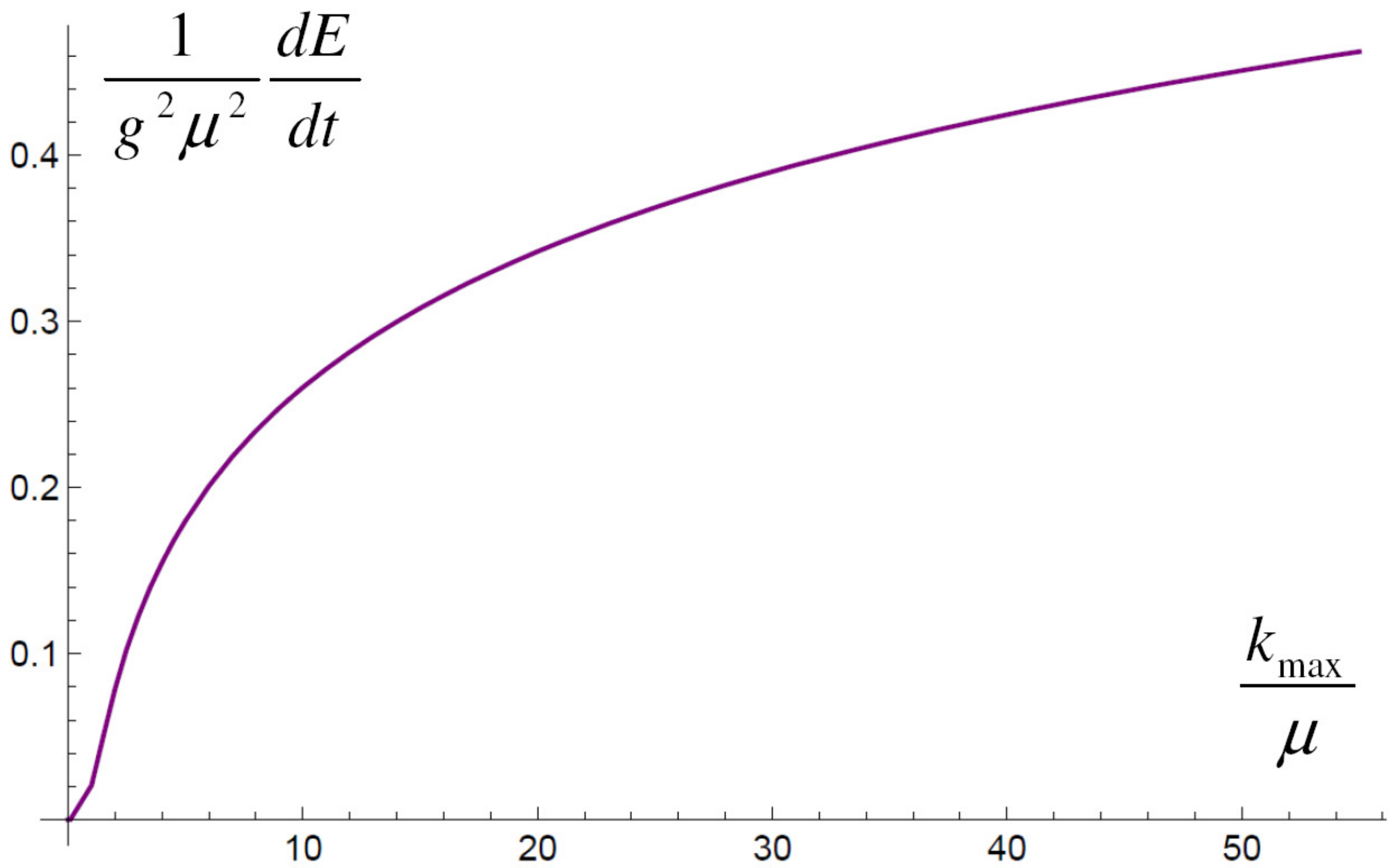}
\caption{The parton energy loss per unit time in equilibrium plasma as a function of $\frac{k_{\rm max}}{\mu}$.}
\label{fig-eq}
\end{figure}

To compare the energy loss in an unstable plasma to that in an equilibrium one, we have computed the integral  in Eq.~(\ref{e-loss-isotropic}) numerically using cylindrical coordinates, which will also be used for the prolate system. Since the integral is known to be logarithmically  divergent,  it has been taken over a finite domain such that  $-k_{\rm max} \le k_L \le k_{\rm max}$ and  $0 \le k_T \le k_{\rm max}$. The energy loss in an equilibrium plasma of massless constituents can be expressed through the Debye mass, which we write as
\be
\label{Debye-mass}
\mu^2 \equiv g^2 \int {d^3p \over (2\pi)^3} \, \frac{f({\bf p})}{|{\bf p}|}.
\ee
We define dimensionless variables by scaling dimensionful quantities by the Debye mass. In Fig.~\ref{fig-eq} we show the energy loss in equilibrium QGP divided by $g^2 \mu^2$ as a function of $\frac{k_{\rm max}}{\mu}$ computed for $g=1$ and $C_R = N_c = 3$. 

\subsection{Unstable plasma}
\label{subsec-unstable}

When the plasma is unstable, the matrix $\Sigma^{-1}(\omega,{\bf k})$ has poles in the upper half-plane of complex $\omega$, and the contributions to the energy loss from these poles grow exponentially in time. The terms in Eq.~(\ref{e-loss-6}) which depend on the initial values of the fields ${\bf D}$ and ${\bf B}$ cannot be neglected, as they exponentially grow  in time. Using the linearized Yang-Mills equations, the initial values ${\bf B}_0$ and ${\bf D}_0$ are expressed through the current and we obtain
\ba
\label{e-loss-unstable}
&& \frac{dE(t)}{dt} = g^2 C_R 
v^i v^l \int_{-\infty +i\sigma}^{\infty +i\sigma}
{d\omega \over 2\pi}
\int {d^3k \over (2\pi)^3}
e^{-i(\omega - \bar{\omega}) t}
(\Sigma^{-1})^{ij}(\omega,{\bf k})
\\ \nn
&&\times
\Big[ 
\frac{\omega \delta^{jl}}{\omega - \bar{\omega}}
-(k^j k^k - {\bf k}^2 \delta^{jk})
(\Sigma^{-1})^{kl}(\bar{\omega},{\bf k}) 
+ \omega \, \bar{\omega} \, \varepsilon^{jk}(\bar{\omega},{\bf k})
(\Sigma^{-1})^{kl}(\bar{\omega},{\bf k})   
 \Big] \,,
\ea
which gives the energy loss of a parton in an unstable quark-gluon plasma. 

\subsection{Inversion of $\Sigma$}
\label{subsec-inv-sigma}

When the anisotropy of the momentum distribution of plasma constituents is controlled by a single (unit) vector ${\bf n}$, it is not difficult to invert the matrix $\Sigma$. Following \cite{Romatschke:2003ms}, we introduce the vector ${\bf n}_T$ defined as
\be
n_T^i \equiv \big(\delta^{ij} - \frac{k^i k^j}{{\bf k}^2}\big) \, n^j
\ee 
and we use the basis of four symmetric tensors
\be
A^{ij}({\bf k}) = \delta^{ij} - \frac{k^i k^j}{{\bf k}^2} ,
\;\;\;\;\;\;
B^{ij}({\bf k}) = \frac{k^i k^j}{{\bf k}^2} ,
\;\;\;\;\;\;
C^{ij}({\bf k},{\bf n}) = \frac{n_T^i n_T^j}{{\bf n}_T^2} ,
\;\;\;\;\;\;
D^{ij}({\bf k},{\bf n}) = 
k^i n_T^j + k^j n_T^i .
\ee
Since the matrix $\Sigma$ is symmetric, it can be decomposed as $\Sigma = a\,A +b\,B +c\,C +d\,D$,  where the coefficients $a$, $b$, $c$ and $d$ are found from the equations
\ba
\label{a-b-c-d}
k^i \Sigma^{ij} k^j = {\bf k}^2 b , \;\;\;\;\;
n_T^i \Sigma^{ij} n_T^j = {\bf n}_T^2 (a + c) , \;\;\;\;\;
n_T^i \Sigma^{ij} k^j = {\bf n}_T^2{\bf k}^2 d , \;\;\;\;\;
{\rm Tr}\Sigma = 2a + b + c .
\ea
The inverse matrix is found to be  
\be
\Sigma^{-1} =
\frac{1}{a} \,A 
+ \frac{-a(a+c)\,B 
+ (- d^2{\bf k}^2{\bf n}_T^2 +bc)\,C
+ad \,D}
{a(d^2{\bf k}^2{\bf n}_T^2-b(a+c))} ,
\ee
and consequently, the poles of the matrix $\Sigma^{-1}$ are given by the following dispersion equations
\be
\label{dis-equations}
a=0 \,, \;\;\;\;\;\;\ 
d^2{\bf k}^2{\bf n}_T^2-b(a+c) = 0 .
\ee

Writing down the energy-loss formula (\ref{e-loss-unstable}) in terms of the projectors $A,B,C,D$ we obtain the form
\ba
\nn
\frac{dE(t)}{dt} 
&=& i g^2 C_R {\bf v}
 \int_{-\infty +i\sigma}^{\infty +i\sigma}
{d\omega \over 2\pi i}
\int {d^3k \over (2\pi)^3}
e^{-i(\omega - \bar{\omega}) t}
\Big( \frac{1}{a} \,A  + \frac{-a(a+c)\,B 
+ (- d^2{\bf k}^2{\bf n}_T^2 +bc)\,C
+ad \,D} {a(d^2{\bf k}^2{\bf n}_T^2-b(a+c))}  \Big)
\\  [2mm] 
\label{e-loss-unstable-A-B-C-D-1}
&\times&
\bigg[ 
\frac{\omega }{\omega - \bar{\omega}} + \frac{\omega}{\bar\omega}
+  \frac{\omega + \bar{\omega}}{\bar\omega} {\bf k}^2
\Big( \frac{1}{\bar{a}} \,A  
+ \frac{  (- \bar{d}^2{\bf k}^2{\bf n}_T^2 +\bar{b}\bar{c})\,C
+\bar{a}\bar{d} \,AD} {\bar{a}(\bar{d}^2{\bf k}^2{\bf n}_T^2-\bar{b}(\bar{a}+\bar{c}))}
\Big) \bigg] {\bf v} ,
\ea
where the coefficients $a,b,c,d$ are computed at $(\omega, {\bf k})$ and the coefficients  $\bar{a},\bar{b},\bar{c},\bar{d}$ at $(\bar{\omega}, {\bf k})$. The formula (\ref{e-loss-unstable-A-B-C-D-1}) will be used in the subsequent section to compute the energy loss in an extremely prolate QGP. 

\section{Extremely prolate plasma}

Anisotropy is a generic feature of the parton momentum distribution in heavy-ion collisions. At the early stage, when partons emerge from the incoming nucleons, the momentum distribution is strongly elongated along the beam - it has a {\em prolate} shape with the average transverse momentum being much smaller than the average longitudinal one. Due to free streaming, the distribution evolves in the local rest frame to a form which is squeezed along the beam - it has {\em oblate} shape with the average transverse momentum being much larger than the average longitudinal one. We consider here the extremely prolate momentum distribution which is either infinitely elongated in the longitudinal direction (defined be the unit vector ${\bf n}$) or infinitely squeezed in the transverse directions. In this case the spectrum of collective excitations can be found analytically. 

\subsection{Momentum distribution}

The extremely prolate momentum distribution can be written as
\be
\label{momentum-dist} 
f({\bf p})= \delta\big({\bf p}^2 - ({\bf p}\cdot{\bf n})^2\big) 
\; h({\bf p}\cdot{\bf n}) \;,
\ee
where $h(x)$ is any positive even function such that $\int d^3 f({\bf p})$ is finite. The integral in Eq. (\ref{Debye-mass}) can be used to define a mass parameter for either isotropic or anisotropic momentum distributions, although in the later case $\mu^{-1}$ cannot be interpreted as the screening length. Applying the prolate distribution a different mass parameter denoted as  $m$ naturally appears. It is related to $\mu$ as $m^2 \equiv \frac{1}{2}\mu^2$.

Since the velocity ${\bf v}$ of a massless parton as given by the distribution (\ref{momentum-dist}) is ${\bf v}={\bf n}$ for ${\bf p} \cdot {\bf n} > 0$ and ${\bf v} = - {\bf n}$ for ${\bf p} \cdot {\bf n} < 0$, the matrix $\Sigma$ defined by Eq.~(\ref{matrix-sigma}) is found to be
\ba
\label{Sigma-final}
\Sigma^{ij}(\omega,{\bf k}) = 
(\omega^2 - m^2 -{\bf k}^2) \delta^{ij} +k^ik^j
&-&\frac{m^2 {\bf k}\cdot {\bf n}}
{\omega^2 - ({\bf k}\cdot {\bf n})^2}
(k^i n^j + n^i k^j)
\\ [2mm] \nn
&-&\frac{m^2 \big(\omega^2 + ({\bf k}\cdot {\bf n})^2\big)
({\bf k}^2 - \omega^2)}
{\big(\omega^2 - ({\bf k}\cdot {\bf n})^2\big)^2}
n^i n^j ,
\ea
and the coefficients $a,b,c,d$ are
\ba
\label{a}
a(\omega,{\bf k}) &=& \omega^2 - m^2 -{\bf k}^2 ,
\\[2mm]
\label{b}
b(\omega,{\bf k}) &=& \omega^2 - m^2 
- \frac{2m^2  ({\bf k}\cdot {\bf n})^2}
{\omega^2 - ({\bf k}\cdot {\bf n})^2}
- \frac{m^2 \big(\omega^2 + ({\bf k}\cdot {\bf n})^2\big)
({\bf k}^2 - \omega^2)}
{\big(\omega^2 - ({\bf k}\cdot {\bf n})^2\big)^2}
\frac{({\bf k}\cdot {\bf n})^2}{{\bf k}^2} ,
\\[2mm]
\label{c}
c(\omega,{\bf k}) &=& 
\frac{m^2(\omega^2+(\mathbf{k}\cdot\mathbf{n})^2)(k^2-\omega^2)}
{(\omega^2-(\mathbf{k}\cdot\mathbf{n})^2)^2}
\left(\frac{(\mathbf{k}\cdot\mathbf{n})^2}{k^2}-1\right) ,
\\[2mm]
\label{d}
d(\omega,{\bf k}) &=& 
- \frac{m^2  ({\bf k}\cdot {\bf n})}
{\omega^2 - ({\bf k}\cdot {\bf n})^2}
- \frac{m^2 \big(\omega^2 + ({\bf k}\cdot {\bf n})^2\big)
({\bf k}^2 - \omega^2)}
{\big(\omega^2 - ({\bf k}\cdot {\bf n})^2\big)^2}
\frac{({\bf k}\cdot {\bf n})}{{\bf k}^2} .
\ea

\begin{figure}[t]
\centering
\includegraphics[width=0.70\textwidth]{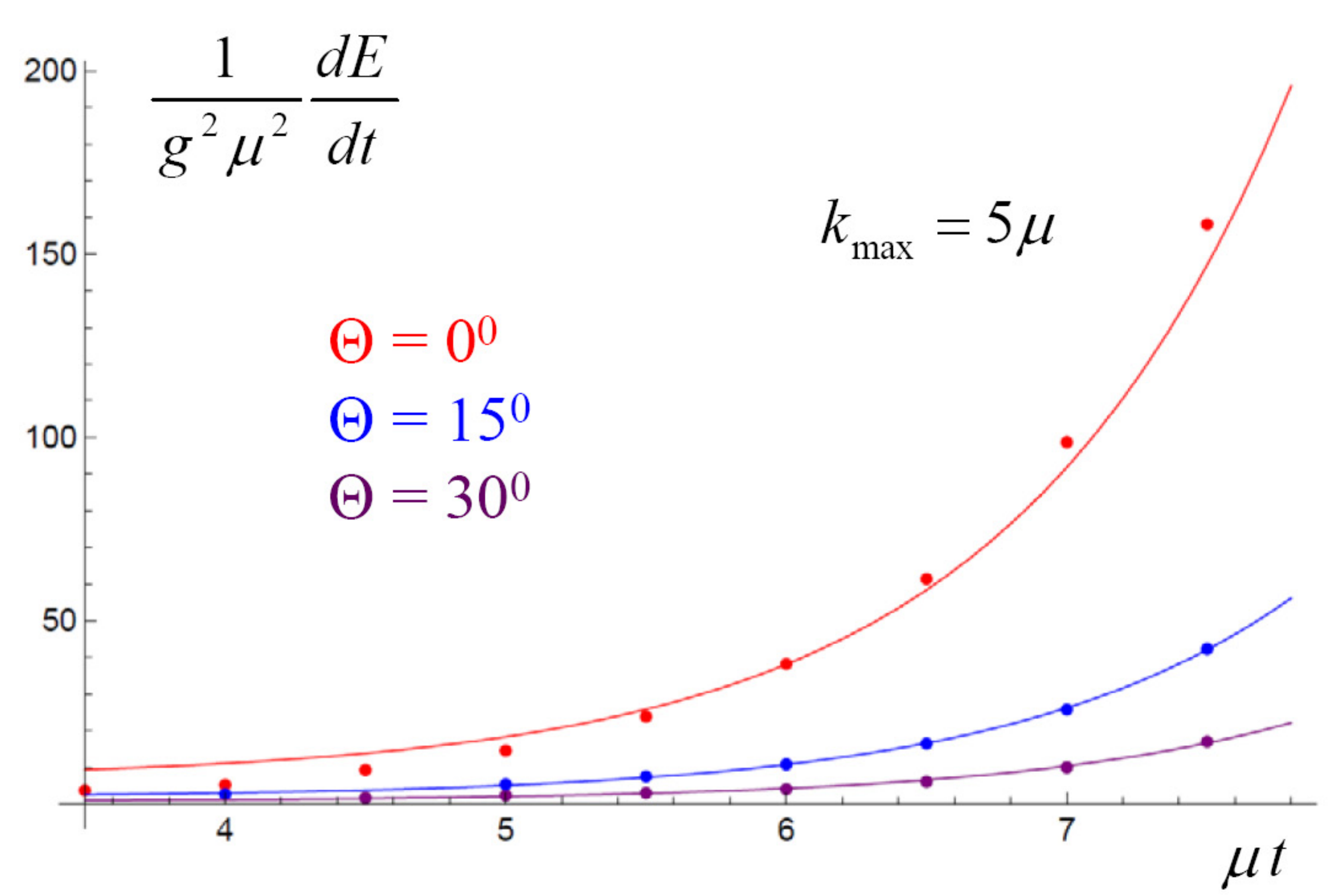}
\caption{The parton energy loss per unit time as a function of time for three angles $\Theta$ between the parton's velocity ${\bf v}$  and the axis $z$. The red points correspond to $\Theta=0$, the blue ones to $\Theta=\pi/12$ and the purple points to $\Theta=\pi/6$. The solid lines represent the exponential fits to the computed points.}
\label{fig-prolate}
\end{figure}

\subsection{Collective excitations}

The dispersion relations are obtained from Eqs. (\ref{dis-equations}) with the coefficients given by Eqs.~(\ref{a}-\ref{d}). The first of these equations provides $\omega_2^2({\bf k}) = m^2 + {\bf k}^2$. Although the second equation looks rather complicated, it has three relatively simple solutions
\ba
\label{general-solution-1}
\omega_1^2({\bf k}) &=& m^2 + ({\bf k}\cdot {\bf n})^2 \;,
\\ [2mm]
\label{general-solution-2}
\omega_{\pm}^2({\bf k}) &=& 
\frac{1}{2}\Big({\bf k}^2 + ({\bf k}\cdot {\bf n})^2
\pm
\sqrt{{\bf k}^4 + ({\bf k}\cdot {\bf n})^4 
+ 4m^2 {\bf k}^2 - 4m^2 ({\bf k}\cdot {\bf n})^2
-2 {\bf k}^2({\bf k}\cdot {\bf n})^2} \; \Big) \;,
\ea
which hold under the condition $\omega^2 \not= ({\bf k}\cdot {\bf n})^2$. The modes $\omega_1$, $\omega_2$ and  $\omega_+$, are always stable. The solution $\omega_-^2$ is negative when $m^2 {\bf k}^2 > m^2 ({\bf k}\cdot {\bf n})^2+ {\bf k}^2({\bf k}\cdot {\bf n})^2$. Writing $\omega_-^2=-\gamma^2$, $0< \gamma \in \mathbb{R}$, the solutions are $\pm i\gamma$. The first is the Weibel unstable mode and the second is an overdamped mode. Denoting the angle between ${\bf k}$ and ${\bf n}$ as $\theta$, the condition for the existence of an instability is $\cos^2 \theta < \frac{m^2}{m^2+{\bf k}^2}$. One can show that for fixed ${\bf k}^2$ the instability growth $\gamma$ is maximal when ${\bf k} \perp {\bf n}$. Collective excitations in the extremely prolate QGP were earlier studied in \cite{Arnold:2003rq} using a method different than ours. 

Using Eqs. (\ref{general-solution-1}) and (\ref{general-solution-2}) one can write
\be
a\big( d^2{\bf k}^2{\bf n}_T^2-b(a+c) \big) = -\frac{\omega^2 \big(\omega^2 - \omega^2_1({\bf k})\big) \big(\omega^2 - \omega^2_2({\bf k})\big) \big(\omega^2 - \omega^2_+({\bf k})\big) \big(\omega^2 - \omega^2_-({\bf k})\big)}{\omega^2 -({\bf k}\cdot {\bf n})^2}\,
\ee 
and from Eq. (\ref{e-loss-unstable-A-B-C-D-1}) one therefore sees that the energy loss in the extremely prolate system is controlled by the double pole at $\omega=0$ and 8 single poles: $\omega = \pm \omega_1$, $\omega = \pm \omega_2$,  $\omega = \pm \omega_+$,  $\omega = \pm \omega_-$.  Since the collective modes are known analytically, the integral over $\omega$ in  Eq.~ (\ref{e-loss-unstable-A-B-C-D-1}) can be computed analytically as well. The remaining integrals are performed numerically.

\subsection{Numerical results}

To compute the integral over ${\bf k}$ in Eq.~ (\ref{e-loss-unstable-A-B-C-D-1}), we use cylindrical coordinates with the $z$ axis  along the vector ${\bf n}$. Since the integral is divergent (as is the case in equilibrium (\ref{e-loss-isotropic})), we choose a finite domain such that  $-k_{\rm max} \le k_L \le k_{\rm max}$ and  $0 \le k_T \le k_{\rm max}$ with $k_{\rm max} = 5 \mu$.  The values of remaining parameters are: $g=1$, $C_R = N_c= 3$. In Fig.~\ref{fig-prolate} we show the parton's energy loss per unit time as a function of time for three different orientations of the parton's velocity ${\bf v}$ with respect to the $z$ axis. The energy loss manifests a strong directional dependence and it exponentially grows in time, which indicates the effect of the unstable modes. After a sufficiently long time, the magnitude of energy loss much exceeds that in equilibrium plasma which, as shown in Fig.~\ref{fig-eq}, equals 0.18 for $k_{\rm max} = 5 \mu$.

\section{Conclusions}

We have developed a formalism where the energy loss of a fast parton in a plasma medium is found as the solution of an initial value problem.  The formalism, which allows one to obtain the energy loss in an unstable plasma, is applied to an extremely prolate quark-gluon plasma with momentum distribution infinitely elongated in the $z$ direction. This system is unstable due to chromomagnetic transverse modes. The energy loss per unit length of a highly energetic parton is not a constant, as in the equilibrium plasma, but it exponentially grows in time and exhibits a strong directional dependence.  The magnitude of the energy loss can much exceed the equilibrium value. 

\section*{Acknowledgment}
This work was partially supported by the Polish National Science Centre under grant 2011/03/B/ST2/00110.

\end{document}